\documentclass{aastex}
\usepackage{spr-astr-addons}
\usepackage{url}
\usepackage{amsfonts}
\usepackage{amssymb,epsfig}
\usepackage{latexsym}
\usepackage{amsmath}
\usepackage{graphicx}

\RequirePackage{color}

\begin{document}

\title{ Holographic dark energy with time varying $c^2$ parameter}

\slugcomment{Not to appear in Nonlearned J., 45.}
\shorttitle{Short article title} \shortauthors{Autors et al.}

\author{M. Malekjani\altaffilmark{1,2}} \and \author{R. Zarei\altaffilmark{1}} \and \author{M. Honari-Jafarpour\altaffilmark{1}}
\altaffiltext{}{E-mail: malekjani@basu.ac.ir.}
\altaffiltext{}{E-mail: M.Honarijafarpour@basu.ac.ir.}
 \altaffiltext{1}{Department of Physics, Faculty
of Science, Bu-Ali Sina University, Hamedan 65178, Iran.}

\altaffiltext{2}{Research Institute for Astronomy $\&$ Astrophysics
of Maragha (RIAAM)- Maragha, Iran, P. O. Box: 55134-441.}

\begin{abstract}
We consider the holographic dark energy model in which the model
parameter $c^2$ evolves slowly with time. First we calculate the
evolution of EoS parameter as well as the deceleration parameter in
this generalized version of holographic dark energy (GHDE).
Depending on the parameter $c^2$, the phantom regime can be achieved
earlier or later compare with original version of holographic dark
energy. The evolution of energy density of GHDE model is
investigated in terms of parameter $c^2$. We also show that the
time-dependency of $c^2$ can effect on the transition epoch from
decelerated phase to accelerated expansion. Finally, we perform the
statefinder diagnostic for GHDE model and show that the evolutionary
trajectories of the model in $s-r$ plane are strongly depend on the
parameter $c^2$.

\end{abstract}

\newpage
\section{Introduction}
Since $1998$, we have a strong belief that our universe experiences
an accelerated expansion. The various astronomical data obtained
from SNe Ia \citep{c1}, WMAP \citep{c2}, SDSS \citep{c3} and X-ray
\citep{c4} experiments support this cosmic acceleration. In the
framework of standard cosmology, a dark energy component with
negative pressure is necessary for this acceleration. The dark
energy scenario is one of the most popular research areas in modern
cosmology. Up to now many theoretical models have been suggested to
interpret the treatment of dark energy. The earliest and simplest
one is the Einstein's cosmological constant with the time -
independent equation of state $w_{\Lambda}=-1$. The cosmological
constant suffers from tow deep theoretical problems namely the
"fine-tuning" and "cosmic coincidence". In addition to cosmological
constant, dynamical dark energy model with time- varying equation of
state have been investigated to interpret the cosmic acceleration.
The scalar field models such as quintessence \citep{c5}, phantom
\citep{c6}, quintom \citep{c7}, K-essence \citep{c8}, tachyon
\citep{c9} and dilaton \citep{c10} together with interacting dark
energy models such as holographic \citep{c11} and agegraphic
\citep{c12} models are the
examples of dynamical dark energy models.\\
The interesting approach of the nature of dark energy is considering
it as an issue of quantum gravity \citep{wit22}. The holographic
dark energy (HDE) is constructed based on the holographic principle
in quantum gravity  \citep{c11}. In holographic principle, a short
distance ultra- violet (UV) cut-off is related to the long distance
infra-red (IR) cut-off, due to the limit set by the formation of a
black hole \citep{c11}. Based on the holographic principle, Cohen et
al.,
 indicated that the zero-point energy of a system with size $L$
 should not be exceed from the mass of black hole with the same
 size, i.e.,
 \begin{equation}
 L^3\rho_d\leq L m_{p}^2
 \end{equation}
 where $L$ is the UV cut-off and $m_{p}$ is the reduced plank mass.
 From the above relation, the energy density of HDE in cosmology can
 be described as
 \begin{equation}
 \rho_d=3c^2m_{p}^2L^{-2}
 \end{equation}
where $c^2$ is a numerical constant of order unity. The HDE model
has been constrained by various astronomical observation
\citep{obs3a,obs1,Wu:2007fs,obs3} and also investigated widely in
the literature \cite{nonflat,holoext}. If we consider $L$ as a
Hubble length $H^{-1}$, in this case, the accelerated expansion of
the universe can not be archived and we get a wrong equation of
state for HDE model \citep{c11}. However, by considering the
interaction between dark matter and dark energy, one can derive the
accelerated expansion of the universe and solve the coincidence
problem for HDE model under Hubble length scale $H^{-1}$
\cite{hublelength}. In the case of particle horizon, the HDE model
can not also obtain the cosmic acceleration \citep{c11}, but this
model with event horizon can derive the accelerated expansion of the
universe \citep{Li}. Therefor we consider the event horizon as an IR
cut-off for HDE model as
\begin{equation}
R_h=a\int_{t}^{\infty}\frac{dt}{a}=a\int_{t}^{\infty}\frac{da^{\prime}}{Ha^{\prime}}
\end{equation}
The coincidence problem has been solved in HDE model with event
horizon \citep{Li}. This model also stable from the view point of
perturbational theory
\citep{HDEstable,casimirHDE,Hogan,jwlee,liwang,hologas}. The
parameter $c^2$ in HDE model has an essential rule in characterizing
the properties of HDE model. For example, the HDE model can behave
as a phantom or quintessence dark energy models at the future for
the values of $c^2$ bigger or smaller than $1$, respectively. In the
standard HDE model the parameter $c^2$ is constant respect with
cosmic time. However there are no strong evidences telling us that
$c$ should be a constant parameter.In general the term $c^2$ can be
assumed as a function of time. By slowly vary function with time,
$\dot{(c^2)}/c^2$ is upper bounded by the Hubble expansion rate,
i.e.,
\begin{equation}
\frac{\dot{(c^2)}}{c^2}\leq H
\end{equation}
 In this case the time scale of the evulsion of $c^2$ is shorter
 than $H^{-1}$ and one can be satisfied to consider the time
 dependency of $c^2$ \citep{radi}. Also, it has been shown that the
 parameter $c^2$ can not be constant for all times during the
 evolution of the universe \citep{radi}. The holographic dark energy
 with time varying $c^2$ at the Hubble length has been studied in
 \citep{duran}. It has been shown that the interacting model of dark
 energy in which the coincidence problem is alleviated can be recast
 as a noninteracting model in which the holographic parameter $c^2$
 evolves slowly with time \citep{duran}. The HDE model with
 time varying $c^2$, the so-called generalized holographic dark
 energy (GHDE), has been constrained by astronomical data \citep{zhang22}. In
GHDE with event horizon, the energy density od dark energy is given
by
 \begin{equation}\label{roghde}
 \rho_d=3c^2(z)m_{p}^2R_h^{-2}
 \end{equation}
It has been shown that the GHDE model can interpret the phantom,
quintessence and cosmological constant models, depending on the
parameter $c(z)$. This generalization has also been done for
holographic ricci dark energy
model \citep{wi22}.\\
Motivated by the above studies, we consider the GHDE model described
by event horizon and obtain the cosmological evolution of the model
in FRW cosmology. Also we apply the statefinder diagnostic for GHDE
model. Since the Hubble parameter, $H=\dot{a}/a$, (first time
derivative) and the deceleration parameter $q=-\ddot{a}H^2/a$
(second time derivative) can not discriminate the model, we need a
higher order of time derivative of scale factor. Sahni et al.
\citep{sah03} and Alam et al. \citep{alamb}, by using the third time
derivative of scale factor, introduced the statefinder pair \{s,r\}
in order to diagnostic the treatment of dark energy models. The
statefinder pair in spatially flat universe is given by
\begin{equation}\label{state1}
r=\frac{\dddot{a}}{aH^3},s=\frac{r-1}{3(q-1/2)}
\end{equation}
The statefinder parameters $s$ and $r$ are the geometrical
parameters, because they only depend on the scale factor. Up to now,
different dark energy models have been investigated from the
viewpoint of statefinder diagnostic. These models have different
evolutionary trajectories in \{s, r\} plane, therefore the
statefinder tool can discriminate these models. The well known
$\Lambda$CDM model corresponds to the fixed point \{s=0,r=1\} in the
$s-r$ plane \citep{sah03}. The distance of the current value of
statefinder pair $\{s_0, r_0\}$ of a given dark energy model from
the fixed point \{s=0,r=1\} is a valuable criterion to examine of
model.\\
Here we list the following dark energy models which have been
studied from the viewpoint of statefinder diagnostic:\\
 the
quintessence DE model \citep{sah03, alamb} , the interacting
quintessence models \citep{zim, zhang055}, the holographic dark
energy models \citep{zhang056, zhang057} , the holographic dark
energy model in non-flat universe \citep{setar}, the phantom model
\citep{chang}, the tachyon \citep{shao}, the generalized chaplygin
gas model \citep{malek11}, the interacting new agegraphic DE model
in flat and non-flat universe  \citep{ zhang058, malek12}, the
agegraphic dark energy model with and without interaction in flat
and non-flat universe \citep{wei077, malek13}, the new holographic
dark energy model \citep{malek20}, the interacting polytropic gas
model \citep{poly99} and the interacting ghost dark energy model \citep{malekg}\\
In this work first we study the cosmological evolution of GHDE model
by calculating the evolution of cosmological quantities EoS and
deceleration parameters. Then we investigate this model from the
viewpoint of statefinder diagnosic.

\section{GHDE model in FRW cosmology}
In the context of flat Friedmann-Robertson-Walker (FRW) cosmology,
the Friedmann equation is given by
\begin{equation}\label{fridt}
H^{2}=\frac{1}{3m_{p}^{2}}(\rho _{m}+\rho _d)
\end{equation}%
where $H$ and $m_p$ are the Hubble parameter and the reduced Planck
mass, respectively.\\
By using the dimensionless energy densities
\begin{equation}\label{denergyt}
\Omega_{m}=\frac{\rho_m}{\rho_c}=\frac{\rho_m}{3M_p^2H^2}, ~~~\\
\Omega_d=\frac{\rho_d}{\rho_c}=\frac{\rho_d}{3M_p^2H^2}~~\\
\end{equation}
 the Friedmann equation (\ref{fridt}) can be
written as
\begin{equation}
\Omega _{m}+\Omega _{\Lambda}=1.  \label{Freqt}
\end{equation}%
 The conservation equations for dark matter and dark energy
are given by
\begin{eqnarray}
\dot{\rho _{m}}+3H\rho _{m}=0, \label{contmt}\\
\dot{\rho _d}+3H(1+w_d)\rho_d=0. \label{contdt}
\end{eqnarray}%
Taking the time derivative of Friedmann equation (\ref{fridt}) and
using (\ref{Freqt}, \ref{contmt}, \ref{contdt}), one can obtain
\begin{equation}\label{hdott}
\frac{\dot{H}}{H^2}=-\frac{3}{2}[1+w_{\Lambda}\Omega_d]
\end{equation}
Taking the time derivative of (\ref{roghde}) and using
(\ref{hdott}), $\dot{R_h}=1+HR_h$,  from (\ref{contdt}), we obtain
the equation of state for GHDE model as follows
\begin{equation}\label{statet1}
w_d=-\frac{1}{3}-\frac{2}{3c}\sqrt{\Omega_d}-\frac{2c^{\prime}}{3c}
\end{equation}
where prime is derivative with respect to $x=\ln{a}$. In terms of
cosmic redshift, we have $d/dx=-(1+z)d/dz$. Taking the derivative
with respect to $x=\ln{a}$, we obtain
\begin{eqnarray}\label{wp}
w_d^{\prime}&=&\frac{-2}{3c}\Big(\frac{\sqrt{\Omega_d}(1-\Omega_d)(1+\frac{2\sqrt{\Omega_d}}{c}+\frac{2c^{\prime}}{c})}{2} \nonumber \\
&&-\frac{c^{\prime}}{c}(c^{\prime}+\sqrt{\Omega_d})+c^{(2)}\Big)
\end{eqnarray}
 where $c^{(2)}=d^2c/dx^2$. Also, taking the time
derivative of $\Omega_d=\rho_d/\rho_c=\frac{1}{H^2R_h^2}$ we obtain
the evolutionary equation for dark energy density for GHDE as
follows
\begin{equation}\label{omegp}
\Omega_d^{\prime}=\Omega_d(1-\Omega_d)(1+\frac{2\sqrt{\Omega_d}}{c}+\frac{2c^{\prime}}{c})
\end{equation}
Using (\ref{hdott})and (\ref{statet1}), the deceleration parameter
$q$ which represents the decelerated or accelerated phase of the
expansion of the universe, for GHDE model can be calculated as
\begin{equation}\label{dec1}
q=-1-\frac{\dot{H}}{H^2}=\frac{1}{2}(1-\Omega_d)-\Omega_d^{3/2}-\frac{c^{\prime}\Omega_d}{c}
\end{equation}
In the limiting case of constant parameter $c$ (i.e.,
$c^{\prime}=0$) all of the above equations reduce to those obtained
for original holographic dark energy (OHDE) model in
\citep{xinzhang}.\\
For complexness, we now derive the statefinder parameters \{s,r\} in
GHDE model. For this aim, by time derivative of (\ref{hdott}), we
first obtain
\begin{equation}\label{ddotht}
\frac{\ddot{H}}{H^3}=\frac{9}{2}(1+w_d\Omega_d)^2-\frac{3}{2}(w_d^{\prime}\Omega_d+w_d\Omega_d^{\prime})
\end{equation}

Inserting (\ref{hdott})and (\ref{ddotht}) in
$r=\ddot{H}/H^3+3\dot{H}/H^2+1$, we obtain the following equation
for the parameter $r$:
\begin{equation}\label{r1}
r=1+\frac{9}{2}w_d\Omega_d(1+w_d\Omega_d)-\frac{3}{2}(w_d^{\prime}\Omega_d+w_d\Omega_d^{\prime})
\end{equation}
The statefinder parameter $s=(1-r)/(9/2+3\dot{H}/H^2)$ is also
obtained as follows
\begin{equation}\label{s1}
s=1+w_d\Omega_d-\frac{1}{3}(\frac{w_d^{\prime}}{w_d}+\frac{\Omega_d^{\prime}}{\Omega_d})
\end{equation}
Putting (\ref{statet1}), (\ref{wp})and (\ref{omegp}) in equations
(\ref{r1}) and (\ref{s1}) yields the following relations for
statefinder parameters of GHDE model:
\begin{eqnarray}\label{r2}
r&=&1-\Big(\frac{3\Omega_d}{2}+\frac{3\Omega_d^{3/2}}{c}+\frac{3c^{\prime}\Omega_d}{c}\Big)\times
\nonumber \\
&&\Big(1-\frac{\Omega_d}{3}-\frac{2\Omega_d^{3/2}}{3c}-\frac{2c^{\prime}\Omega_d}{c}\Big)\\
\nonumber
 &&+\Big(\frac{\Omega_d^{3/2}(1-\Omega_d)(1+\frac{2\sqrt{\Omega_d}}{c}+\frac{2c^{\prime}}{c})}{2c}
 \nonumber \\
 &&-\frac{c^{\prime}\Omega_d(c^{\prime}+\sqrt{\Omega_d})}{c^2}+\frac{c^{(2)}\Omega_d}{c}\Big) \nonumber \\
 &&+\Big((\frac{\Omega_d}{2}+\frac{\Omega_d^{3/2}}{c}+\frac{c^{\prime}\Omega_d}{c})(1-\Omega_d)(1+\frac{2\sqrt{\Omega_d}}{c}+\frac{2c^{\prime}}{c})\Big)
 \nonumber
\end{eqnarray}
and

\begin{eqnarray}\label{s2}
s&=&1-(\frac{\Omega_d}{3}+\frac{2\Omega_d^{3/2}}{3c}+\frac{2c^{\prime}\Omega_d}{3c})
\nonumber \\
&&-\frac{1}{3}(1-\Omega_d)(1+\frac{3\sqrt{\Omega_d}}{c}+\frac{2c^{\prime}}{c})\\
\nonumber
 &&+\frac{2}{c(1+\frac{2\sqrt{\Omega_d}}{c}+\frac{2c^{\prime}}{c})}(\frac{c^{\prime}(c^{\prime}+\sqrt{\Omega_d})}{c}+\frac{c^{(2)}}{3})
\end{eqnarray}
where $c^{(2)}=d^2c/dx^2$.\\
In the next section we give a numerical description of the evolution
of GHDE model by solving the equations
(\ref{statet1},\ref{omegp},\ref{wp},\ref{dec1},\ref{r2},\ref{s2}).
Here we consider the model parameter $c(z)$ of GHDE as a function of
redshift as follows
\begin{equation}\label{c1z}
c(z)=c_0+c_1\frac{z}{1+z}
\end{equation}
The above choice for $c(z)$ is inspired by the parameterizations
known as Chevallier-Polarski-Linder (CPL) \citep{cpl}. At the early
time ($z\rightarrow \infty$), we have $c\rightarrow c_0+c_1$ and at
the present time ($z\rightarrow 0$), $c\rightarrow c_0$. Therefore
the model parameter $c$ varies smoothly from $c_0+c_1$ to $c_0$ from
past to present. By the above choice, the first and second
derivative of $c$, i.e., $c^{\prime}$ and $c^{(2)}$ are
\begin{equation}\label{c2z}
c^{\prime}=-c_1/1+z, ~~ c^{(2)}=-c_1/1+z,
\end{equation}
respectively. Assuming the positive energy density of GHDE model at
any time yields the following conditions for $c_0$ and $c_1$:
\begin{equation}\label{c3z}
c_0>0, ~~ c_0+c_1>0
\end{equation}

\section{Numerical results}
Here we calculate the evolutionary behavior of GHDE model in FRW
cosmology. We first obtain the evolution of EoS parameter as well as
the deceleration parameter. Then we perform the statefinder
diagnosis and $w-w^{\prime}$ analysis for this model. In numerical
procedure we set $\Omega_m=0.30$ and $\Omega_d=0.70$.
\subsection{EoS parameter}
By solving (\ref{statet1}) and using (\ref{c1z}, \ref{c2z}) for
different model parameters $c_0$ and $c_1$, we show the evolution of
EoS parameter of GHDE as a function of redshift in Fig.(1). In upper
panel we fix the parameter $c_1=0.10$ and vary the parameter
$c_0=0.25,0.5,0.75$ corresponds to the solid-blue, dashed-black and
dotted-dashed-red curves, respectively. Here we see that the GHDE
model enters the phantom regime without a need for interaction
between dark matter and dark energy. Also, it is worthwhile to
mention that the GHDE model crosses that phantom line ($w_d=-1$)
from up ($w_d>-1$) to below ($w_d<-1$). This behavior of GHDE model
is in agreement with recent observations \citep{obse1}. By
increasing the parameter $c_0$ can be achieved later. In lower
panel, by fixing $c_0=0.70$, we vary the parameter
$c_1=-0.10,0.00,0.10$ corresponds to dashed-black, solid-blue and
dotted-dashed-red curves, respectively. The solid-blue curve
indicates the original holographic dark energy model (OHDE). One can
conclude that for $c_1<0$ the GHDE model can cross the phantom line
earlier and for $c_1>0$ cross the phantom line later compare with
OHDE model. It should be noted that the above illustrative values
for $c_0$ and $c_1$ should satisfy the conditions in (\ref{c3z}).
\subsection{energy density}
Here we calculate the evolution of energy density of GHDE model as a
function of redshift parameter from the early time up to late time
by solving equation (\ref{omegp}). The evolution of parameters $c$
and $c^{\prime}$ are given by (\ref{c1z}) and (\ref{c2z}),
respectively. In Fig.(2), we plot the evolution of energy density
$\Omega_d$ versus of redshift for different values of model
parameters $c_0$ and $c_1$. We see that at the early times
$\Omega_d\rightarrow 0$ and at the late times $\Omega_d\rightarrow
1$, means the dark energy dominated universe at the late time. In
upper panel by fixing $c_1=0.10$ the parameter $c_0$ is varied as
illustrative values $0.25, 0.50 , 0.75$ corresponding to solid-blue,
dashed-black and dotted-dashed-red curves, respectively. We see that
in the past times the dark energy becomes more dominant for larger
values of $c_0$ and at the late times the dark energy dominated
universe can be archived sooner for lower values. In lower panel by
fixing $c_0=0.70$ the parameter $c_1$ is varied as illustrative
values $-0.10, 0.00 , 0.10$ corresponding to dashed-black,
solid-blue and dotted-dashed-red curves, respectively. It has been
seen that the dark energy becomes more dominant for positive values
of $c_1$ and less dominant for negative values compare with OHDE
model.
\subsection{deceleration parameter}
Here we study the expansion phase of the universe by calculating the
evolution of deceleration parameter $q$ in GHDE model. By solving
equation (\ref{dec1})  and using (\ref{omegp}), we plot the
evolution of $q$ versus redshift parameter $z$ in Fig.(3). In both
panels we see that the parameter $q$ start from $q=0.50$,
representing the $CDM$ model at the early time. Then the parameter
$q$ becomes negative, representing the accelerated expansion phase
of the universe at recent epochs. Therefore the GHDE model can
interpret the decelerated phase of the expansion of the universe at
the early times and accelerated phase later. In upper panel we fix
the parameter $c_1=0.1$ and vary the parameter $c_0=0.25, 0.50,
0.75$ corresponding to solid-blue, dashed-black and
dotted-dashed-red curves, respectively. By increasing $c_0$, the
accelerated expansion can be achieved sooner. In rlower panel, we
fix $c_0=0.70$ and vary $c_1=-0.10, 0.0, 0.10$, corresponding to
dashed-black, solid-blue and dotted-dashed- red curves,
respectively. The solid-blue curve indicate the OHDE model. We see
that negative values of $c_1$ result the larger accelerated
expansion at the present time and positive values of $c_1$ obtain
the smaller accelerated expansion, compare with standard OHDE model.
\subsection{statefinder diagnosis}
The statefinder pair \{s, r\} for GHDE model is given by (\ref{r2})
and (\ref{s2}). In statefnder plane, the horizontal axis is defined
by the parameter $s$ and vertical axis by the parameter $r$. In
Fig.(4), by solving (\ref{s2}) and (\ref{r2}) and using (\ref{c1z},
\ref{c2z}, \ref{c3z}), we obtain the evolutionary trajectories of
GHDE model in $s-r$ plane. In both panels, by expanding the
universe, the evolutionary trajectories evolve from right to left.
The parameter $r$ increases and the parameter $s$ decreases. The
trajectories cross the $\Lambda$CDM fixed point $\{s=0, r=0\}$ at
the middle of evolution. In upper panel we fix the parameter
$c_1=0.10$ and vary the parameter $c_0=0.25, 0.50, 0.75$
corresponding to the solid-blue, dashed-black and dotted-dashed- red
curves, respectively. We see that different values of model
parameter $c_0$ result different trajectories in $s-r$ plane.
Therefore the GHDE model in $s-r$ plane is discriminated for
different values of model parameter $c_0$. The colored circles on
the curves represent the today's value of statefinder parameters
$\{s_0, r_0\}$ of the model. We also see that for larger values of
$c_0$, the distance of $\{s_0, r_0\}$ from the $\Lambda$CDM fixed
point is shorter. In lower panel, the parameter $c_0$ is fixed by
$c_0=0.70$ and the parameter $c_1$ is varied by $c_1=-0.10, 0.00,
0.10$, respectively, corresponding to the dashed-black, solid-blue
and dotted-dashed- red curves. Same as upper panel the GHDE model
mimics the $\Lambda$CDM model at the middle of evolution. The GHDE
model can be discriminated by model parameter $c_1$ in $s-r$ plane.
Different values of $c_1$ result different evolutionary
trajectories. The solid blue curve indicate the OHDE model. We see
that for positive values of $c_1$, the distance of $\{s_0, r_0\}$
from $\Lambda$CDM fixed point is shorter and for negative values of
$c_1$ is longer than standard OHDE model.

\newpage

\section{conclusion}
Summarizing this work, we studied the new version of holographic
dark energy model, the so-called generalized holographic dark energy
(GHDE), in which the model parameter $c^2$ is considered as a
time-varying function. Here we considered the CPL parameterizations
in which $c(z)=c_0+c_1z/(1+z)$ \citep{cpl}. We first investigated
the cosmological evolution of GHDE model by calculating the
evolution of EoS and deceleration parameters. We showed that for
negative values of $c_1$ the phantom regime can be achieved sooner
and for positive values later compare with original holographic
model (OHDE). In agreement with recent observation \citep{obse1}, we
show that the phantom line is crossed from quintessence regime
($w_d>-1$) to phantom regime ($w_d<-1$). The evolution of dark
energy density in terms of model parameter $c^2$ has been
investigated. We showed that the dark energy becomes more dominant
for positive values of $c_1$ and less dominant for negative values
compare with OHDE model. It has been shown that the transition from
decelerated to the accelerated expansion depends on the time-varying
function $c^2(z)$. Increasing the parameter $c_0$ causes that the
transition tacks place sooner. Also positive values of $c_1$ result
larger accelerated expansion and negative values obtain smaller
accelerated expansion compare with OHDE. Eventually we performed the
statefinder diagnostic tool in this model. Different values of $c_0$
and $c_1$ give different evolutionary trajectories for GHDE model in
$s-r$ plane. Hence the GHDE model can be discriminated by parameter
$c^2(z)$. Since the standard $\Lambda$CDM model is still a standard
model of dark energy, therefore a distance of present value
$\{s_0,r_0\}$ from $\Lambda$CDM fixed point $\{s_0=0,r_0=1\}$ is
valuable criterion to examine a given dark energy model in $s-r$
plane. The distance of $\{s_0,r_0\}$ from $\{s_0=0,r_0=1\}$ is
shorter for $c_1>0$ and longer for $c_1<0$ in comparison with OHDE
model ($c_1=0$). Increasing the parameter $c_0$ yields the shorter
distance of $\{s_0,r_0\}$ from $\Lambda$CDM fixed point
$\{s_0=0,r_0=1\}$.

\noindent{\bf{ Acknowledgements}}\\
This work has been supported financially by Research Institute for
Astronomy $\&$ Astrophysics of Maragha (RIAAM) under research
project 1/2782-54.

\newpage

\begin{figure}[!htb]
\includegraphics[width=8cm]{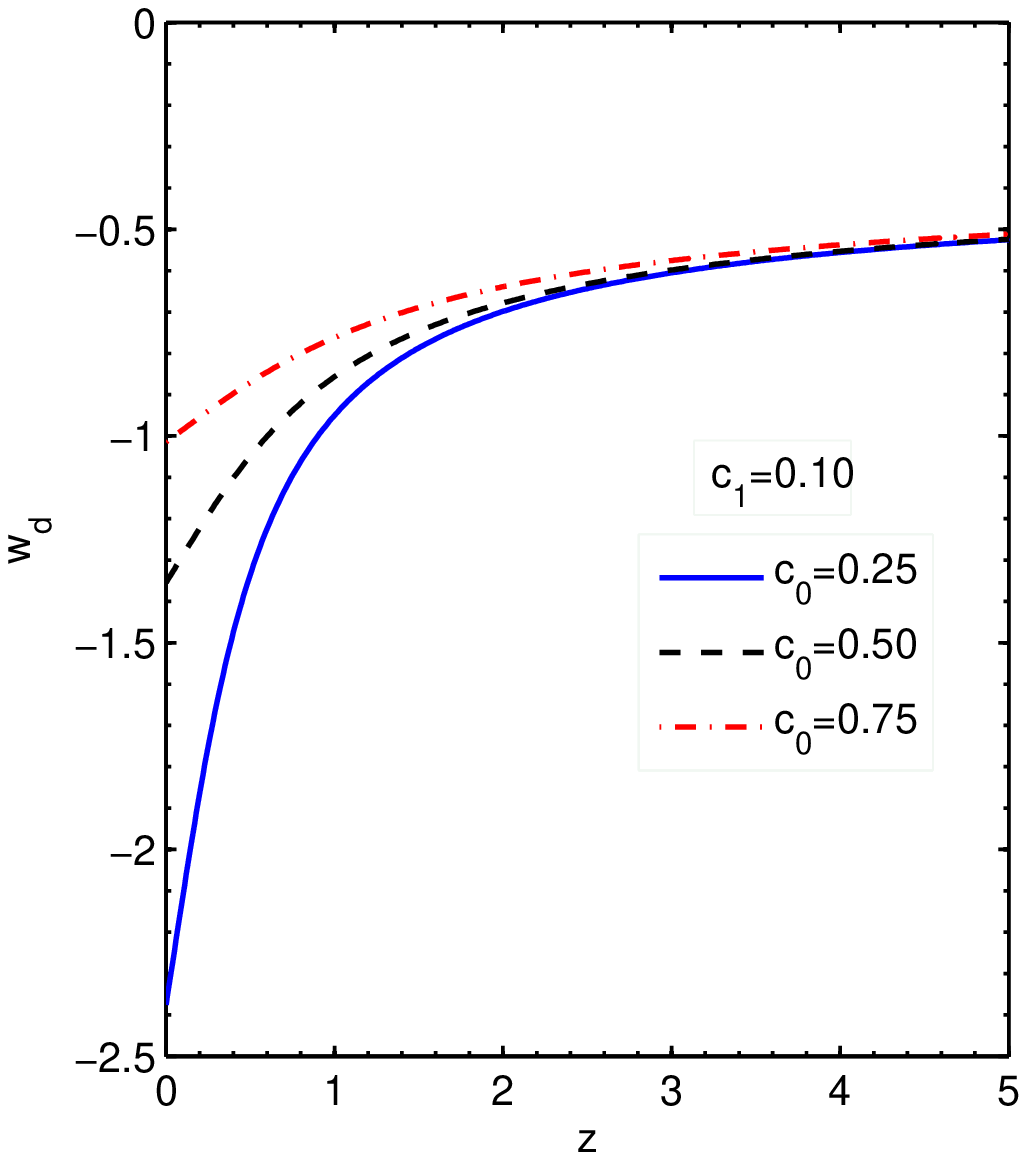} \includegraphics[width=8cm]{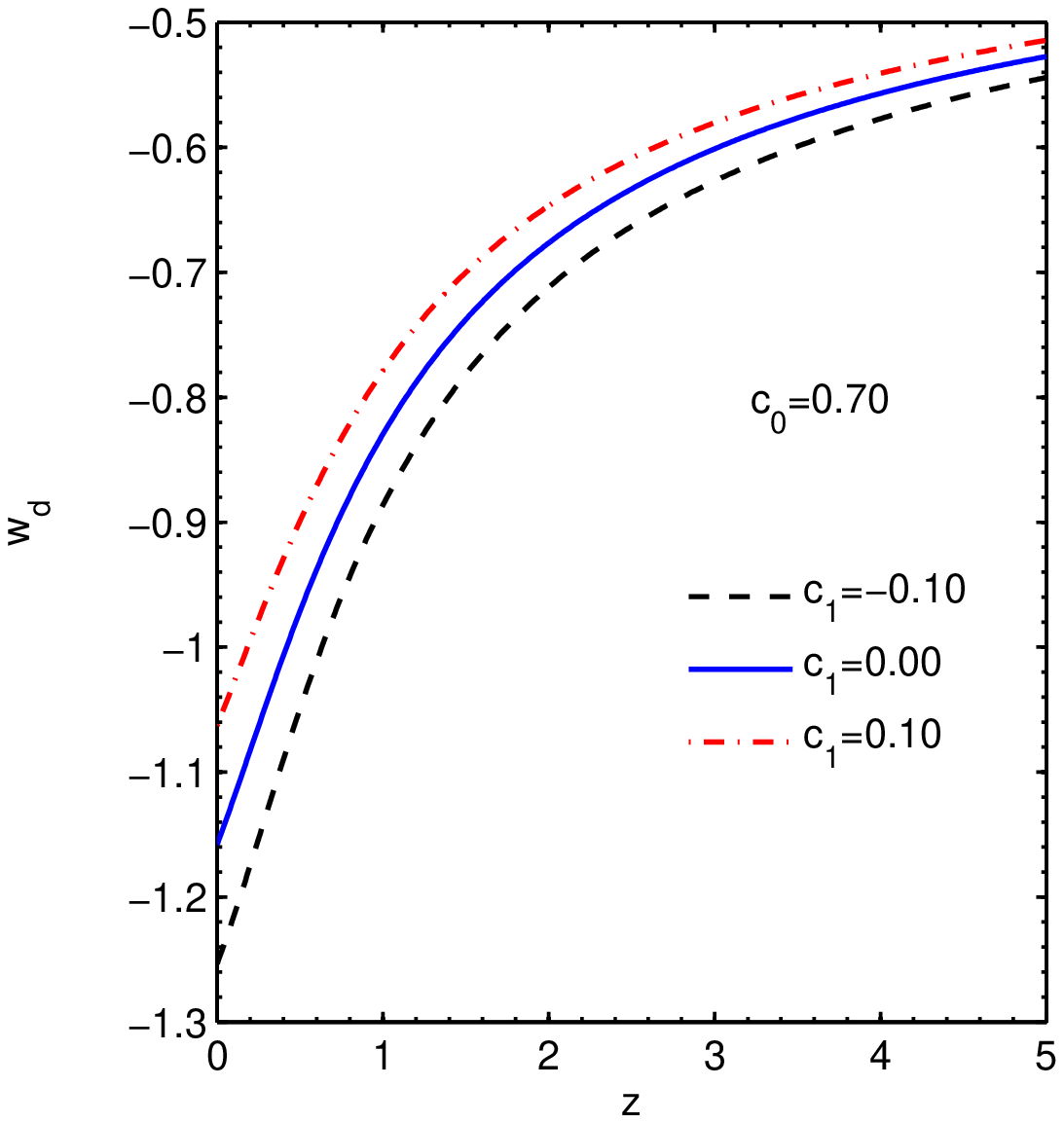}
\caption{The evolution of EoS parameter of GHDE model, $w_{d}$,
versus redshift parameter $z$ for different values of model
parameters $c_0$ and $c_1$. In upper panel, by fixing $c_1$, we vary
$c_0$ as indicated in legend. In lower panel, by fixing $c_0$ we
vary the parameter $c_1$ as described in legend.}
\end{figure}

\begin{figure}[!htb]
\includegraphics[width=8cm]{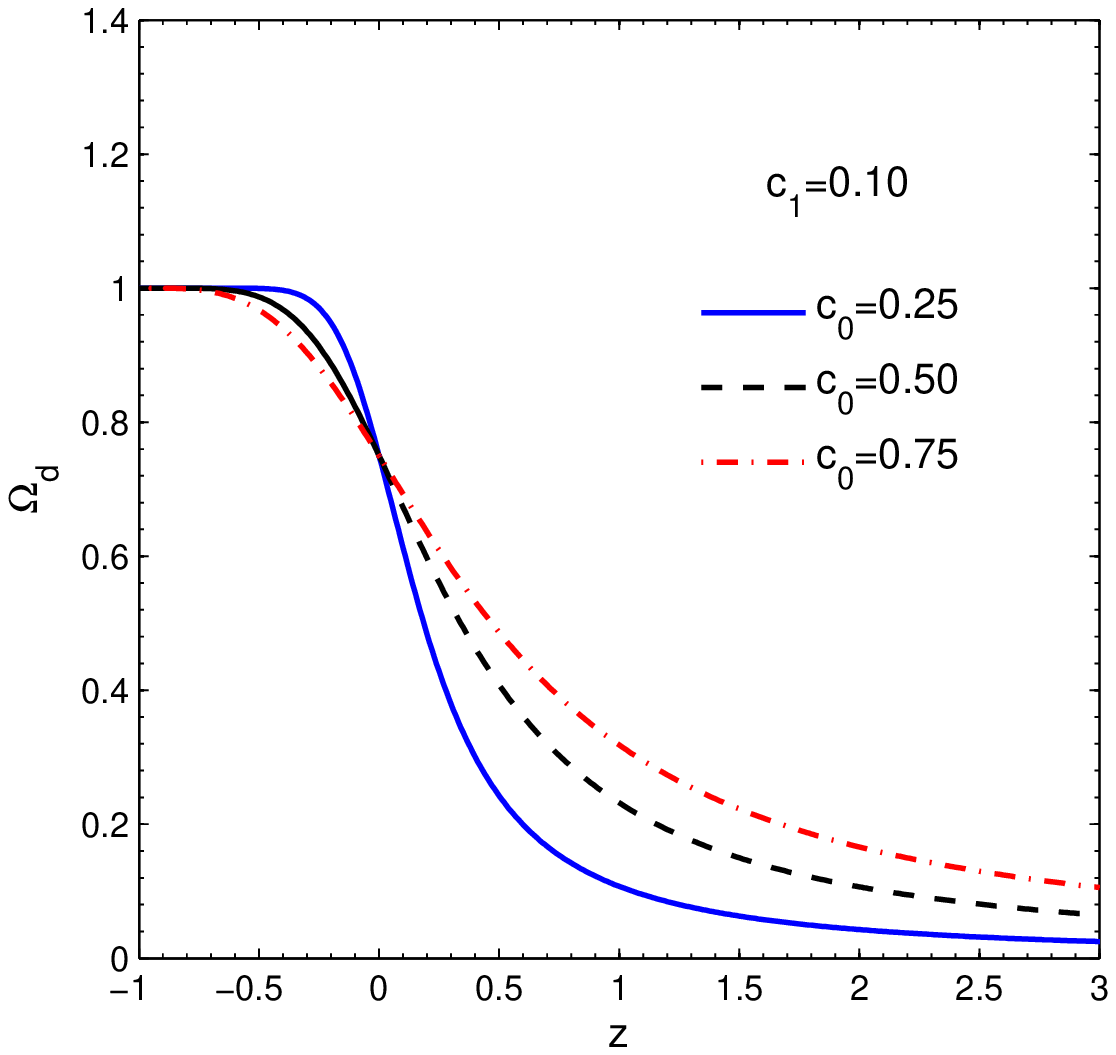} \includegraphics[width=8cm]{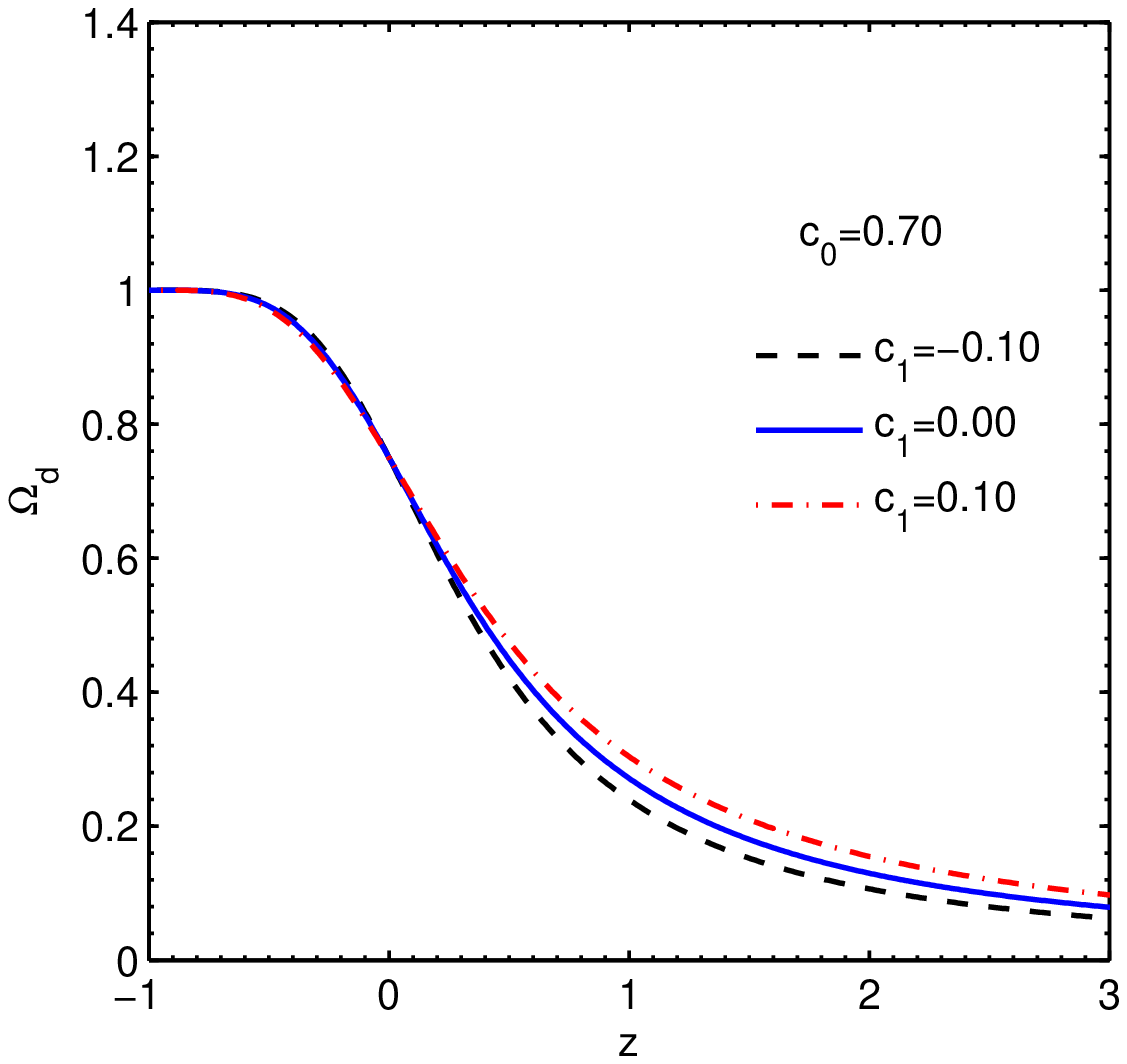}
\caption{The evolution of dark energy density of GHDE model as a
function of redshift $z$. In upper panel, by fixing $c_1$, we vary
$c_0$ as indicated in legend. In lower panel, by fixing $c_0$ we
vary the parameter $c_1$ as described in legend.}
\end{figure}

\begin{figure}[!htb]
\includegraphics[width=8cm]{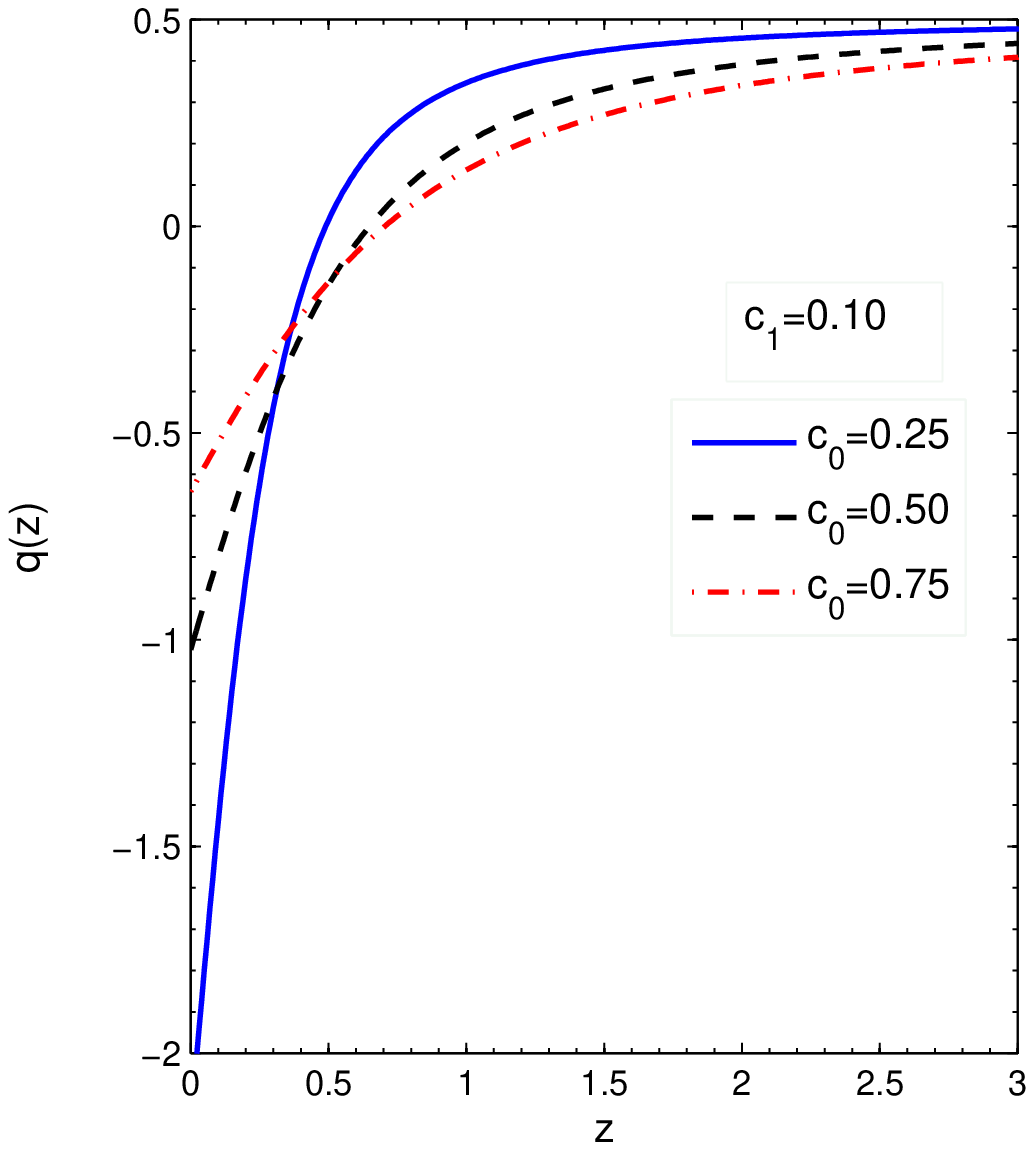} \includegraphics[width=8cm]{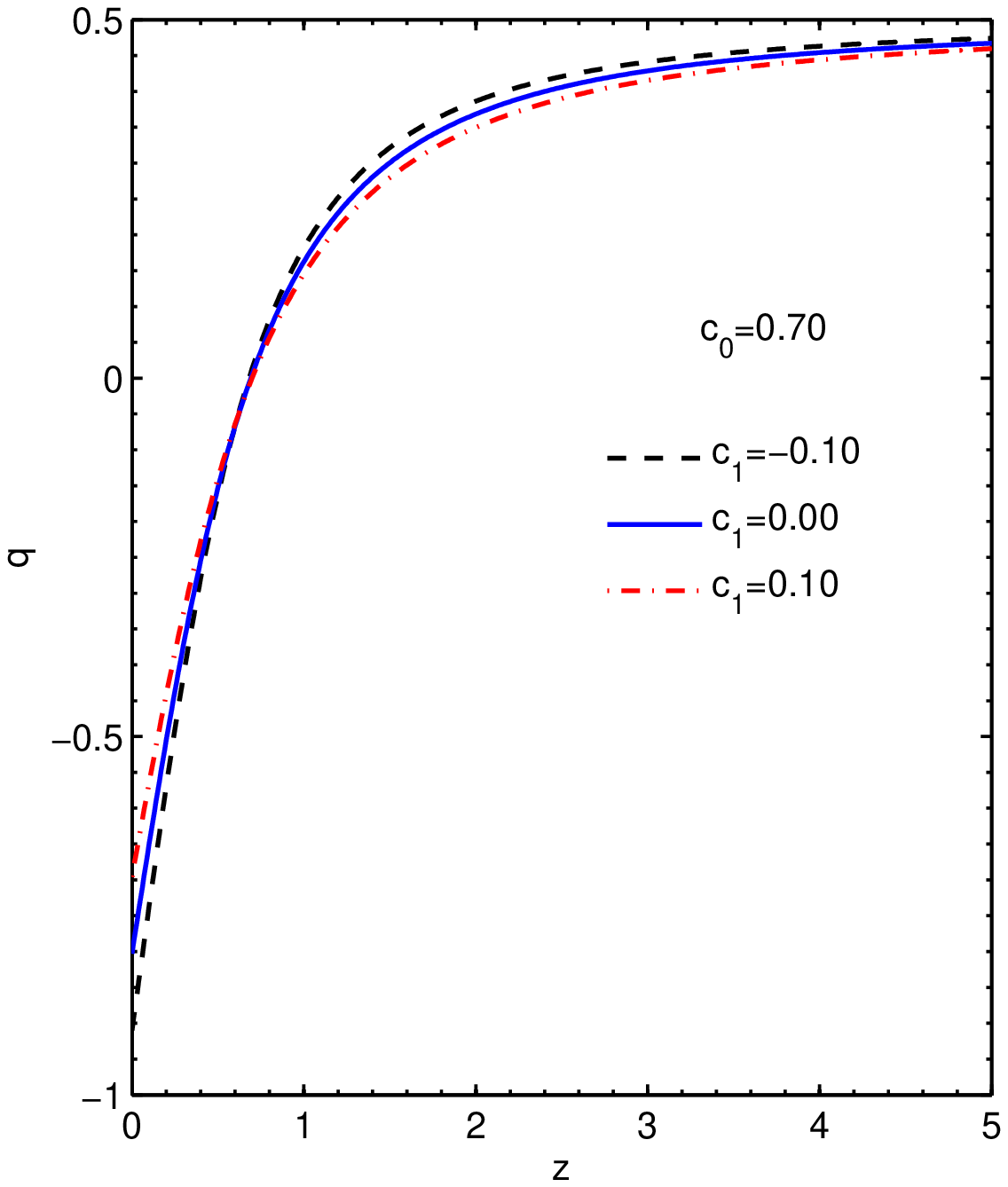}
\caption{The evolution of deceleration parameter $q$ in the context
of GHDE model as a function of redshift parameter $z$ for different
illustrative values of model parameters $c_0$ and $c_1$. In upper
panel, by fixing $c_1$, we vary $c_0$ as indicated in legend. In
lower panel, by fixing $c_0$ we vary the parameter $c_1$ as
described in legend.}
\end{figure}

\begin{figure}[!htb]
\includegraphics[width=8cm]{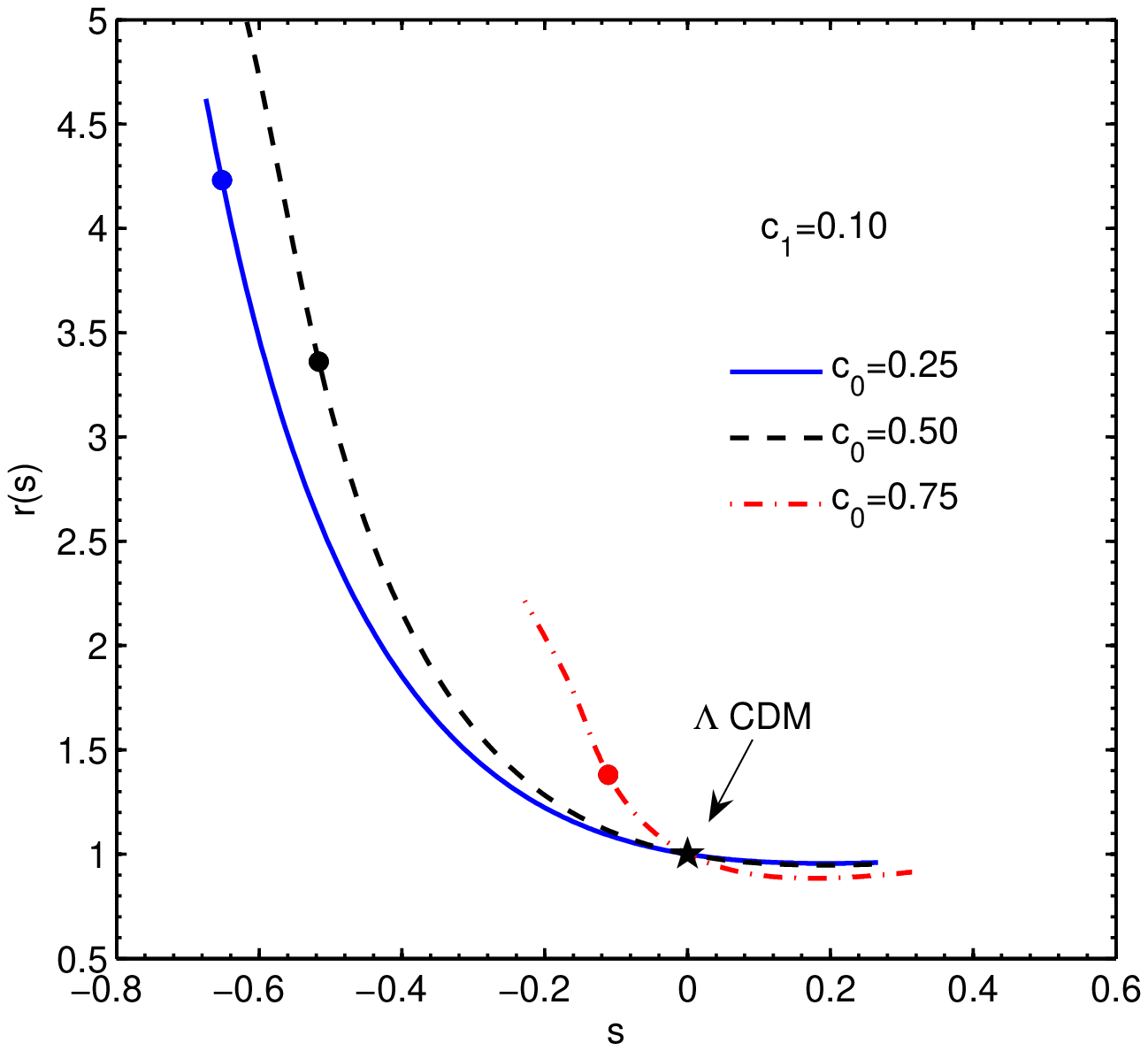} \includegraphics[width=8cm]{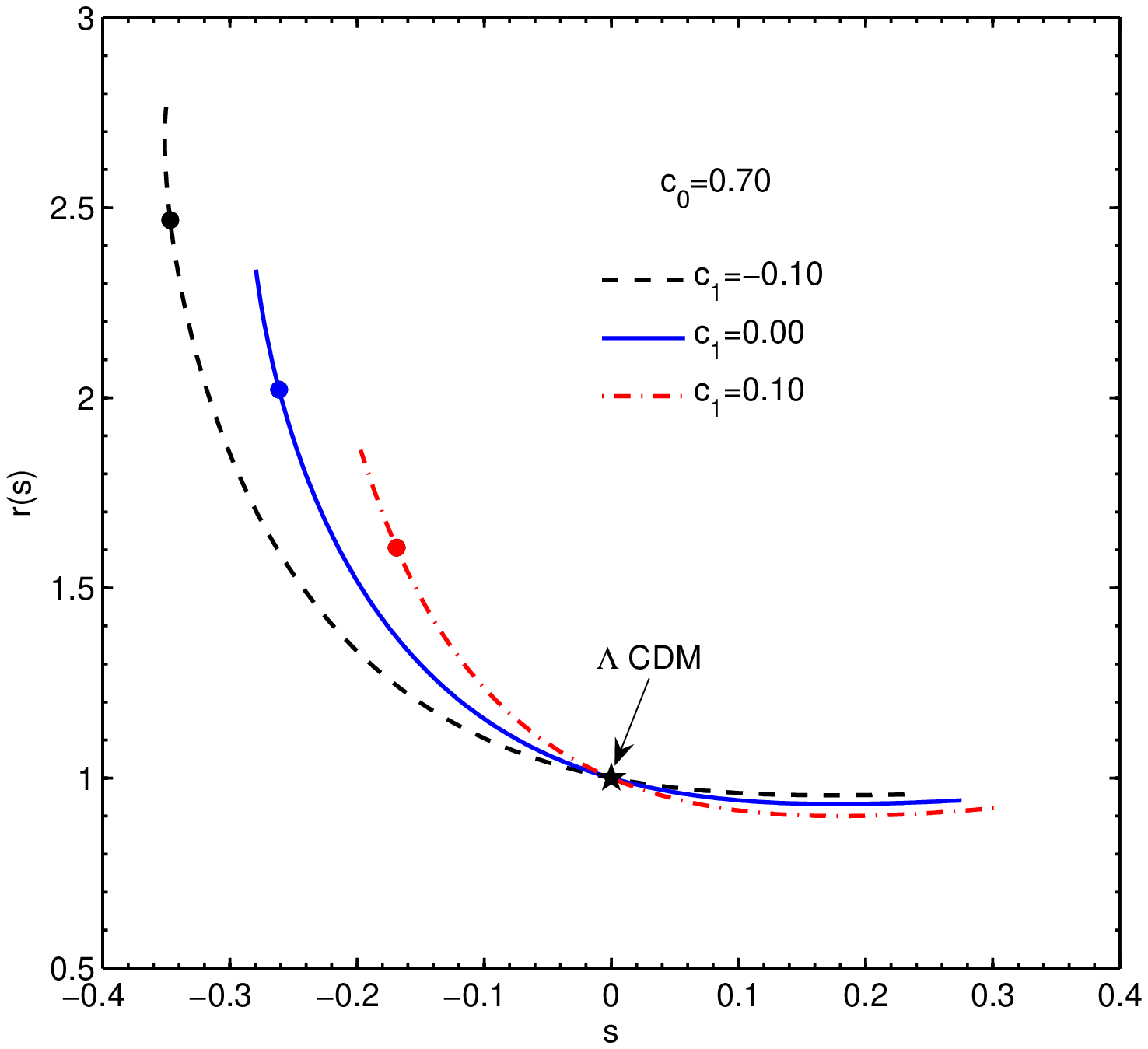}
\caption{The evolutionary trajectories of GHDE model in $s-r$ plane
for different values of model parameters $c_0$ and $c_1$ as
indicated in legend.}
\end{figure}

\end{document}